\documentstyle[12pt,epsf]{article}
\newcommand{\abstracts}[1]{{
\centering{\begin{minipage}{12.5truecm}
\normalsize\baselineskip=15pt
\centerline{\footnotesize ABSTRACT}\vspace*{0.3cm}
\parindent=20pt #1
\end{minipage}}\par}}

\newcommand{\beqn}{\begin{eqnarray}}
\newcommand{\eeqn}{\end{eqnarray}}
\newcommand{\eq}[1]{(\ref{#1})}
\newcommand{\beq}{\begin{equation}}
\newcommand{\eeq}{\end{equation}}

\newcommand{\dual}{\mbox{}^{\ast}}

\makeatletter
\def\leqq{\mathrel{\mathpalette\gl@align<}}
\def\geqq{\mathrel{\mathpalette\gl@align>}}
\def\gl@align#1#2{\lower.6ex\vbox{\baselineskip\z@skip\lineskip\z@
    \ialign{$\m@th#1\hfil##\hfil$\crcr#2\crcr=\crcr}}}
\makeatother

\makeatletter
\def\sileqq{\mathrel{\mathpalette\gs@align<}}
\def\sigeqq{\mathrel{\mathpalette\gs@align>}}
\def\gs@align#1#2{\lower.6ex\vbox{\baselineskip\z@skip\lineskip\z@
    \ialign{$\m@th#1\hfil##\hfil$\crcr#2\crcr\sim\crcr}}}
\makeatother

\def\bbbone{{\mathchoice {\rm 1\mskip-4mu l} {\rm 1\mskip-4mu l}
{\rm 1\mskip-4.5mu l} {\rm 1\mskip-5mu l}}}
\def\bbbc{{\mathchoice {\setbox0=\hbox{$\displaystyle\rm C$}\hbox{\hbox
to0pt{\kern0.4\wd0\vrule height0.9\ht0\hss}\box0}}
{\setbox0=\hbox{$\textstyle\rm C$}\hbox{\hbox
to0pt{\kern0.4\wd0\vrule height0.9\ht0\hss}\box0}}
{\setbox0=\hbox{$\scriptstyle\rm C$}\hbox{\hbox
to0pt{\kern0.4\wd0\vrule height0.9\ht0\hss}\box0}}
{\setbox0=\hbox{$\scriptscriptstyle\rm C$}\hbox{\hbox
to0pt{\kern0.4\wd0\vrule height0.9\ht0\hss}\box0}}}}
\def\bbbe{{\mathchoice {\setbox0=\hbox{\smalletextfont e}\hbox{\raise
0.1\ht0\hbox to0pt{\kern0.4\wd0\vrule width0.3pt
height0.7\ht0\hss}\box0}}
{\setbox0=\hbox{\smalletextfont e}\hbox{\raise
0.1\ht0\hbox to0pt{\kern0.4\wd0\vrule width0.3pt
height0.7\ht0\hss}\box0}}
{\setbox0=\hbox{\smallescriptfont e}\hbox{\raise
0.1\ht0\hbox to0pt{\kern0.5\wd0\vrule width0.2pt
height0.7\ht0\hss}\box0}}
{\setbox0=\hbox{\smallescriptscriptfont e}\hbox{\raise
0.1\ht0\hbox to0pt{\kern0.4\wd0\vrule width0.2pt
height0.7\ht0\hss}\box0}}}}

\begin{document}

~
\vspace{-1cm}
\begin{flushright}
{\large ITEP-TH-12/98}

\vspace{0.2cm}
{\large KANAZAWA-98-03}

\vspace{0.2cm}
{\large ZIF-MS-30/98}

\vspace{0.3cm}
{\sl May 18, 1998, rev. June 9}
\end{flushright}

\begin{center}

{\baselineskip=16pt
{\Large \bf Embedded Topological Defects
in Hot Electroweak}\\
\vspace{0.3cm}
{\Large \bf Theory: a Lattice Study}

\vspace{1cm}

{\large
M.~N.~Chernodub\footnote{chernodub@vxitep.itep.ru}$^{\! ,a}$,
F.~V.~Gubarev\footnote{Fedor.Gubarev@itep.ru}$^{\! ,a}$,
E.--M.~Ilgenfritz\footnote{ilgenfri@hep.s.kanazawa-u.ac.jp}$^{\! ,b}$\\
and A.~Schiller\footnote{schiller@tph204.physik.uni-leipzig.de}$^{\! ,
c,d}$}\\

\vspace{.5cm}
{ \it

$^a$ ITEP, B.Cheremushkinskaya 25, Moscow, 117259, Russia

\vspace{0.3cm}

$^b$ Institute for Theoretical Physics, Kanazawa University,\\
Kanazawa 920-1192, Japan

\vspace{0.3cm}

$^c$ Institut f\"ur Theoretische Physik, Universit\"at  Leipzig,\\
D-04109 Leipzig, Germany

\vspace{0.3cm}

$^d$ Zentrum f\"ur interdisziplin\"are Forschung, Universit\"at
Bielefeld, \\
D-33615 Bielefeld, Germany
}}
\end{center}
\vspace{5mm}

\abstracts{
We study the properties of Nambu monopoles and $Z$--vortices in the $3D$ 
lattice $SU(2)$ Higgs theory which represents the Standard Model at high 
temperature. We show that the densities of the Nambu monopoles and the 
$Z$--vortices are $O(1)$ in the symmetric phase and generically small in the 
Higgs phase. Near to the critical Higgs mass and in the vicinity of the phase 
transition the densities are no more negligible in the broken phase.
The percolation probability of the $Z$--vortex lines is found as a new disorder
parameter for this phase transition. We conclude that the transition to the 
symmetric phase is accompanied by $Z$--vortex condensation. Simulations 
comparing elementary and extended vortices and monopoles at different $\beta_G$
values, aiming to show that the density of vortices and monopoles of fixed 
physical size might have a well-defined continuum limit, gives encouraging but
so far inconclusive results.}

\baselineskip=14pt
\setcounter{footnote}{0}
\renewcommand{\thefootnote}{\alph{footnote}}

\section{Introduction}

This paper is devoted to the study of topological properties of thermal states 
in the standard model of electroweak interactions. Although the standard model 
does not possess {\it topologically stable} monopole- and vortex-like
structures, one can define so-called {\it embedded} topological 
defects~\cite{VaBa69,BaVaBu94}: they are known under the names of Nambu 
monopoles~\cite{Na77} and $Z$--strings~\cite{Na77,Ma83}. We perform the 
investigation of the equilibrium thermodynamics of embedded topological defects
in the framework of dimensional reduction which is expected to be reliable
to describe the physics near to the transition temperatures
(of a truly first order phase transition or a continuous crossover)
for Higgs masses between $30$ and $240$ GeV.

Embedded topological defects play a crucial role in some
electroweak baryogenesis scenarios. For example, one of the scenarios
is based on the decay of an electroweak string network as the
Universe cools down. According to this mechanism, long electroweak
strings decay into smaller, twisted and linked string loops which
carry non-zero baryon number. This could then explain the emergence
of non-zero baryon number in the Universe~\cite{StringScenario}.
Another example is the bound state picture of a sphaleron:  a Nambu
monopole--anti-monopole pair connected by a segment of
$Z$--string~\cite{BaVaBu94,Hi94,ChGuIl97} builds the unstable (saddle
point) sphaleron field configuration~\cite{Ma83,DaHaNe74}
in the broken phase.  After the electroweak transition is completed, this 
type of electroweak configuration is the only transition state available 
between vacua of different Chern--Simons number.
Thermal activation would allow to wash-out the baryon number
asymmetry~\cite{SphRev} and therefore has to be suppressed.
In a recent paper \cite{ChGuIl97} we have analyzed
classical lattice sphalerons (as obtained and discussed in
Ref. \cite{vanbaal}) and have found evidence for the monopole--anti-monopole 
bound state picture of these configurations.

In the present publication, we apply the tools for detecting Nambu
monopoles and $Z$--vortices developed in Ref. \cite{ChGuIl97} in order to study
numerically the formation of these defects in thermal equilibrium
states of the electroweak theory. We measure the densities of Nambu
monopoles and of $Z$--strings across the phase transition. In an attempt to 
substantiate the picture of $Z$--vortex condensation we also study the 
percolation probability of the $Z$--vortex line network near to the phase 
transition temperature. Recently, in a different context, in 
Ref.~\cite{antunes} convincing evidence has been provided that string 
percolation is a good disorder parameter for a phase transition in field 
theory.

Due to the smallness of the Weinberg angle $\theta_W$ the $U(1)$
component of the electroweak group $SU(2) \times U(1)$ can have only
little effect on the properties of the embedded defects.
Thus the dynamics of the topological defects can
be studied restricting ourselves to the $SU(2)$ Higgs sector.
The emerging effective $3D$ $SU(2)$ Higgs lattice model\footnote{Besides
symmetric and Higgs phases, the lattice $SU(2)$ Higgs model in $4D$ has also
the confinement phase which is located at sufficiently small
$\beta_G$ (which has, however, nothing to do with the
electroweak theory in the continuum). One can show analytically that the
transition between the symmetric and the confinement phase in that
model is accompanied by Nambu monopole condensation \cite{Ch97}.
This observation leads us to conjecture that topological defects may
be relevant for the phase transitions also in the physically relevant
region of the parameters of the $SU(2)$ Higgs model,
which is studied here in its dimensionally reduced variant.}
allows to take into account effects of chiral fermions on physical parameters 
perturbatively \cite{thermal,wirNP97}. However, as in our study of the 
structure of the electroweak sphaleron one may expect that the influence of 
the standard model fermions on the embedded defects is small~\cite{SphRev}.

The structure of the paper is as follows. The effective three-dimensional model
is shortly recalled in Section~2. In this Section we also give the lattice 
definitions of the embedded topological defects according to 
Ref.~\cite{ChGuIl97} and generalize them to cover the case of extended defects,
too. In Section~3 we present the numerical results on the density of 
elementary (size $a$) monopoles and vortices and on the percolation 
probability of the corresponding $Z$--vortex lines. In Section~4 we describe 
a first step to extend this analysis to extended vortices and monopoles of 
equal physical size living on finer lattices in order to see whether the 
density discontinuities at the transition temperature possess a continuum 
limit. Section~5 contains a discussion of our results and conclusions.

\section{Embedded Defects in Lattice $SU(2)$ Higgs Model}

We study the lattice $3D$ $SU(2)$ Higgs model with the following action:
\beqn
S &=& \beta_G \sum_p \Bigl(1 - \frac{1}{2} \mbox{\rm Tr} U_p \Bigr)
- \beta_H \sum_{x,\mu} \frac{1}{2} \mbox{\rm Tr}
(\Phi_x^+ U_{x, \mu} \Phi_{x + \hat\mu})
 + \sum_x \biggl( \rho_x^2 + \beta_R (\rho_x^2-1)^2 \biggr)\,,
\eeqn
where the summation is taken over plaquettes $p$, sites $x$ and
links $l=\{x,\mu\}$. The action contains three parameters: the gauge
coupling $\beta_G$, the lattice Higgs self-coupling $\beta_R$ and
the hopping parameter $\beta_H$.  The gauge fields are represented by
unitary $2 \times 2$ link matrices $U_{x,\mu}$ and $U_p$ denotes
the $SU(2)$ plaquette matrix. The Higgs fields are parametrized as
follows: $\Phi_x = \rho_x V_x$, where $\rho_x^2= \frac12 \mbox{\rm
Tr}(\Phi_x^+\Phi_x)$ is the Higgs modulus squared,
and $V_x$ an element of the group $SU(2)$.
The lattice parameters are related to the continuum parameters of the
$3D$ superrenormalizable $SU(2)$ Higgs model $g_3$, $\lambda_3$ and
$m_3(\mu_3=g_3^2)$ as given {\it e.g.} in \cite{wirNP97}. As in \cite{wirNP97} 
a parameter $M_H^*$ is used (approximately equal to the zero temperature 
physical Higgs mass) which expresses the Higgs self-coupling as follows
\beqn
\beta_R=\frac{\lambda_3}{g_3^2}
\frac{\beta_H^2}{\beta_G} = \frac{1}{8}
{\left(\frac{M_H^*}{80\ {\mbox {GeV}}}\right)}^2
\frac{\beta_H^2}{\beta_G}\,.
\label{MH*}
\eeqn
The lattice coupling $\beta_G$ and continuum coupling $g^2_3$ are related by
\beqn
\beta_G = \frac{4}{a g^2_3}\,, \label{betaG}
\eeqn
with $a$ being the lattice spacing. We shall study the theory at different 
gauge couplings $\beta_G$ in order to understand the qualitative behavior of 
the lattice embedded defects towards the continuum limit, {\it i.e.} when the 
lattice becomes fine enough to resolve the structure of defects which then 
have to be described as extended defects.

Let us first define elementary topological defects.
The gauge invariant and quantized lattice definition of the Nambu
monopole is closely related to the definition in the continuum
theory~\cite{ChGuIl97}. First we define a composite adjoint unit
vector field $n_x = n^a_x \, \sigma^a$, $n^a_x = - (\Phi^+_x
\sigma^a \Phi_x) \slash (\Phi^+_x \Phi_x )$
with $\sigma^a$ being the Pauli matrix. The field $n_x$ plays a
role similar to the direction of the adjoint Higgs field in the
definition of the 't~Hooft--Polyakov monopole~\cite{tHPo74} in the
Georgi--Glashow model. Then we define
the gauge invariant flux plaquette~${\bar \theta}_p$:
\beqn
{\bar \theta}_p (U,n) = \arg \Bigl( {\mathrm {Tr}}
 \left[(\bbbone + n_x) V_{x,\mu} V_{x +\hat\mu,\nu}
 V^+_{x + \hat\nu,\mu} V^+_{x,\nu} \right]\Bigr)\,.
 \label{AP}
\eeqn
Here
\beqn
V_{x,\mu}(U,n) = U_{x,\mu} + n_x U_{x,\mu} n_{x + \hat\mu}
\eeqn
has been used to define a particular projection of the gauge field.
The sense of it becomes more transparent if we consider
the unitary gauge where
we have $\Phi_x={(0,\phi)}^T$, $n^a_x
\equiv \delta^{a3}$. In this gauge the phases $\theta^u_l = \arg
U^{11}_l$ behave as a compact Abelian field
with respect to the residual Abelian gauge transformations
$\Omega^{abel}_x = e^{i \sigma_3 \, \alpha_x}$, $\alpha_x \in [0,2
\pi)$ which leave the unitary gauge condition intact. The Nambu monopoles 
are the topological defects in this Abelian field. The gauge invariant 
plaquette function~\eq{AP} coincides with the standard Abelian plaquette 
formed out of the fields~$\theta^u_l$~\cite{ChGuIl97}. In this case, the 
Nambu  monopole charges can be defined using the standard construction:
\beqn
j_c = - \frac{1}{2\pi} \sum_{p \in \partial c}
{\bar \theta}_p\,, \quad
{\bar \theta}_p = \left( \theta_p - 2 \pi m_p \right) \in [-\pi,\pi)\,.
\label{j_N}
\eeqn
In the gauge independent definition (\ref{AP}) the flux immediately
gets values in the required interval $[-\pi,\pi)$ and is
independent of the choice of the reference point $x$
on the plaquette, due to the property
$n_x V_{x,\mu} = V_{x,\mu} n_{x+\hat\mu}$.

The $Z$--string~\cite{Ma83,Na77} corresponds to the Abrikosov--Nielsen--Olesen 
vortex solution~\cite{ANO} embedded~\cite{VaBa69,BaVaBu94} into the 
electroweak theory. The $Z$--vorticity number corresponding to the plaquette 
$p=\{x,\mu\nu\}$ is defined~\cite{ChGuIl97} similar to the construction of 
the Abelian vortices in an Abelian Higgs theory:
\beqn
\sigma_p = \frac{1}{2\pi} \Bigl( \chi_p - {\bar \theta}_p \Bigr) \,,
\label{SigmaN}
\eeqn
where ${\bar \theta}_p$ has been given in eq.~\eq{AP}, and
$\chi_{x,\mu\nu} =
\chi_{x,\mu} + \chi_{x +\hat\mu,\nu} - \chi_{x + \hat\nu,\mu} -
\chi_{x,\nu}$ is the plaquette variable
formed out of Abelian links
\beqn
\chi_{x,\mu} =
\arg\left(\Phi^+_x V_{x,\mu} \Phi_{x + \hat\mu}\right) \,.
\eeqn

The $Z$--vortex is formed by the links $l=\{x,\rho\}$  of
the dual lattice carrying non-vanishing vorticity $\dual \sigma_l$. These links
are dual to the plaquettes $p=\{x,\mu\nu\}$ with non-zero
vortex number~\eq{SigmaN} and $\dual \sigma_{x,\rho} =
\varepsilon_{\rho\mu\nu} \sigma_{x,\mu\nu} \slash 2$. One can
show that $Z$--vortices begin and end on the Nambu (anti-) monopoles:
$\sum^3_{\mu=1} (\dual \sigma_{x-\hat\mu,\mu} - \dual \sigma_{x,\mu})
= \dual j_x$.

In order to understand the behavior of the embedded defects towards
the continuum limit we have made measurements of densities of Nambu
monopoles and $Z$--vortices {\it just of one definite physical size}
which amounts to different multiples of $a$ on different lattices.
We follow the change from the broken to the symmetric phase at various
gauge couplings, $\beta_G=8,16,24$, {\it i.e.} with a lattice step
size $a$ getting smaller with increasing $\beta_G$. In principle, starting 
from a lattice simulated at various $\beta_G$, we could perform different 
factor $k$ real space renormalization group transformations of the 
gauge-Higgs field configurations in order to measure properties of defects 
of many different sizes $k a$ objects in a manner described above.
Instead of doing this, as a first step, we use here the notion of 
{\it extended} topological objects on the lattice as invented in 
Ref.~\cite{IvPoPo90}. An extended monopole (vortex) of
physical size $k a$ is defined on $k^3$ cubes ($k^2$ plaquettes,
respectively).\footnote{In the context of Abelian projections of
pure Yang--Mills theory this construction is known as type-I
extended monopoles \cite{IvPoPo90}.
So-called type-II monopoles have been defined in the same context.
These are defined by adding the charges (currents) of elementary
monopoles inside the extended cube $c(n)$, and type-II vortices can
be constructed similarly. We have also tried this construction for
our model and found that here this block spin transformation does not
seem to lead to a well-defined continuum limit for extended defects
contrary to the case of $SU(2)$ gluodynamics where the type-II
Abelian monopoles behave according to the renormalization
group~\cite{ShibaSuzuki}.
In the limit $a \to 0$ the detection of size $a$ topological defects
with the definition given above is afflicted with increasing lattice
artifacts.}
The charge of monopoles $j_{c(k)}$ on bigger $k^3$ cubes $c(k)$ is
constructed analogously to that of the elementary
monopole, eq. \eq{j_N}, with the elementary $1\times 1$ plaquettes
in terms of $V_{x,\mu}$ replaced by $n \times n$ Wilson loops 
(extended plaquettes) denoted as $p(n)$.

First, the projected links $V_{x,\mu}(U,n)$ are defined as usual,
then extended $V$-links
\beqn
V^{(k)}_{X,\mu}=V_{X,\mu}V_{X +\hat\mu,\mu}...V_{X +(k-1) \hat\mu,\mu}
\eeqn
are formed where $X$ and $X + k \hat\mu$ denote points of the decimated 
lattice, which inherit the property
$n_{X} V^{(k)}_{X,\mu} = V^{(k)}_{X,\mu} n_{X + k \hat\mu}$.
Finally, the phase angles of length-$k$ extended Higgs operators are defined as
\beqn
\chi_{l(k)}=\chi^{(k)}_{X,\mu} =
\arg\left(\Phi^+_X V^{(k)}_{X,\mu}
\Phi_{X + k \hat\mu}\right) \,.
\eeqn

With the gauge invariant flux~${\bar \theta}_{p(k)}$ for the extended
plaquettes defined analogously to (\ref{AP}) in terms of the extended 
$V$-links and with the extended plaquette angles $\chi_{p(k)}$ formed
out of the extended links angles $\chi_{l(k)}$, the definitions of
extended monopoles and vortices can be taken over from 
eqs.~(\ref{j_N},\ref{SigmaN})
\beqn
j_{c(k)} = - \frac{1}{2\pi} \sum_{p(k) \in \partial c(k)}
{\bar \theta}_{p(k)}\,, \quad
\sigma_{p(k)} = \frac{1}{2\pi} \Bigl( \chi_{p(k)}
- {\bar \theta}_{p(k)} \Bigr) \,.
\label{jk_N}
\eeqn
According to eq. \eq{betaG} the physical size of the $k^3$ monopoles
(or the $k^2$ vortices) in simulations done at $\beta_G = k\beta^{(0)}_G$
should be equal for all $k$. Roughly the same physical
objects are hopefully described as elementary monopoles and
vortices at $\beta^{(0)}_G$, too.

\section{Scanning the Transition with Size $a$ Defects}

First, the formation of
{\it elementary} Nambu monopoles and $Z$--vortices has been scanned across
the electroweak phase transition. Monte Carlo simulations have
been performed on cubic lattices of size $L^3=16^3$ at $\beta_G=12$
for self-couplings $\lambda_3$ corresponding to
$M_H^*=30$~GeV and $M_H^*=70$~GeV, see
eq.~(\ref{MH*}).  We varied the parameter $\beta_H$ in order to
locate the phase transition for given $M_H^*$ and $\beta_G$. For the
update we used the algorithms described in Ref.~\cite{wirNP97} which
combine Gaussian heat bath updates for the gauge and Higgs fields
with several reflections for the fields to reduce the
autocorrelations.
Usually, the first order nature of the thermal transition is detected
by a two-peak histogram in the volume average of $\rho^2$.
The average over all Monte Carlo configurations
$\langle \rho^2 \rangle$ denotes the quadratic Higgs condensate, the
discontinuity (for infinite volume) of which is related to the latent
heat \cite{wirNP97}.

Here we are interested in the behavior of the lattice densities
$\rho_m$ of Nambu monopoles and $\rho_v$ of $Z$--vortices and want to measure
the percolation probability for the $Z$--vortex lines near to the
phase transition. For each lattice configuration, the
densities $\rho_m$ and $\rho_v$ are given by
\beqn
 \rho_m = \frac{1}{L^3} \sum\limits_c |j_c|\,,
 \qquad
 \rho_v = \frac{1}{3 L^3} \sum\limits_p |\sigma_p|\,,
\eeqn
where $c$ and $p$ refers to elementary cubes and plaquettes.
The monopole charge $j_c$ and the $Z$--vorticity $\sigma_p$ are
defined in eq.~\eq{j_N} and eq.~\eq{SigmaN}, respectively.

The percolation probability of the vortex lines $\dual \sigma$ is defined 
with the help of the following two-point function~\cite{PoPoYu91}:
\beqn
C(r) & = & {\left(\sum\limits_{x,y,i}
\delta_{x \in \dual \sigma^{(i)}} \,\delta_{y \in \dual \sigma^{(i)}}
\cdot \delta\Bigl(|x-y|-r\Bigr) \right)} \cdot {\left(
\sum\limits_{x,y} \delta\Bigl(|x-y|-r\Bigr) \right)}^{-1}\,,
\eeqn
where the summation is taken over all points $x$, $y$ of the dual
lattice and over all connected clusters of vortex lines $\dual
\sigma^{(i)}$ ($i$ labels distinct vortex clusters).
The Euclidean distance between two points $x$ and $y$ is denoted as $|x-y|$.
The notation $x \in \dual \sigma^{(i)}$ means that the vortex world line 
cluster $\dual \sigma^{(i)}$ passes through the point $x$.

Then the percolation probability is
\beqn
C & = & \lim_{r \to \infty} C(r)\,.
\eeqn
This formula corresponds to
the thermodynamical limit. In this case the function $C(r)$ can be
fitted as $C(r) = C + C_0 r^{-\alpha} e^{-m r}$, with $C$,
$C_0$, $\alpha$ and $m$ being fitting parameters. In our finite
volume we find numerically that $m \sim a^{-1}$ in the studied region of
the phase diagram. Therefore we can be sure that
finite size corrections to $C$ are exponentially suppressed.
If $C$ does not vanish, {\it i.e.} if two infinitely separated points with
non-zero probability belong to the same connected cluster of
vortex lines, the vacuum is populated by {\it infinitely long} vortex
lines. In other words, an effective theory of these vortices
would have an action smaller than the entropy per length of the vortex
trajectory. This case implies the existence of a vortex condensate.
If $C$ turns to zero the vortex condensate vanishes~\cite{PoPoYu91}.

In Figure~1(a) we show the ensemble averages of densities, 
$\langle \rho_m \rangle$ of Nambu monopoles and $\langle \rho_v \rangle$ of
$Z$--vortices {\it vs.} hopping parameter $\beta_H$
for Higgs mass $M_H^*= 30$~GeV at gauge coupling $\beta_G = 12$.
At this Higgs mass the phase transition is of strongly first
order \cite{wirPL95}.
It is seen that the averages of both densities are $O(1)$ in the symmetric
phase and they sharply drop at the phase transition to zero densities in the 
Higgs phase. In Figure~1(b) the
percolation probability $C$ for the same parameters is shown
to have a small discontinuity at the transition. Deeper into the
symmetric phase the condensate (percolation probability) grows while it
is strictly vanishing in the broken phase.

For comparison, Figure~2(a) presents the Monte Carlo averages
$\langle \rho_m \rangle$ and $\langle \rho_v \rangle$ for the Higgs
mass $M_H^*=70$ GeV at $\beta_G = 12$. This Higgs mass parameter is
very near to the critical Higgs mass. Now, the average densities
of Nambu monopoles and $Z$--strings do not vanish immediately
after the phase transition towards the broken phase: both densities are
non-zero at temperatures $T \sileqq T_c$.  Still, the corresponding
vortices {\it do not percolate}. The percolation probability $C$ for
the $Z$--vortex trajectories behaves as a disorder
parameter also in {\it this} case when the high-temperature electroweak
phase transition is already very weakly first order \cite{wirNP97}.
It turns to zero exactly at $\beta_{Hc}$ as can be seen in Figure~2(b) but
a continuous behavior cannot be excluded from these data.

At the larger Higgs mass $M_H^*=70$ GeV, just for the value of
the gauge coupling $\beta_G=12$ studied before, we have also tried to 
understand how the infinite volume limit is approached. We show in Fig.~3(a)
and Fig.~3(b) histograms of the Nambu monopole density $\rho_m$ and
of the $Z$--vortices $\rho_v$, respectively, for lattice sizes
$16^3$, $24^3$, $32^3$ and $40^3$.  Even near to the end of the first
order phase transition, we see that not exceedingly large lattices
are necessary to see the two-state signal in the densities and to
measure the (finite volume) discontinuity of the densities, such that its
infinite volume limit can be obtained. In the present state of our analysis
we cannot yet present corresponding histograms of the percolation
probability which would allow to tell whether the weakly first order phase
transition is still accompanied by a discontinuous change of the vortex
condensate. Independent of this, one may say that
an infinite percolating network is breaking into finite pieces
at the transition temperature while the correlation length stays finite.

\section{Is There a Continuum Limit ?}

In Ref.~\cite{Laine98} lattice vortices have been studied in the
non-compact version of the $3D$ Abelian Higgs model. The $3D$ Abelian Higgs
model has become a model system for many phase transitions, covering
the electroweak and the QCD deconfinement transition, superconductivity
as well as cosmic strings.  From their attempt to define a well-defined
continuum limit of the vortex density two lessons can be drawn. First,
the lattice density might be afflicted by regularization artifacts such
that, generally speaking, discontinuities at the phase transition are
physically better defined. Second, one should carefully associate the density
of defects with (various) physical scales and try to find the continuum limit 
of these. At least at the strongly first order phase transition at small Higgs 
mass the first problem seems to be not to so severe. But in order to get a 
physically well-defined vortex and monopole density, the extrapolation to the 
continuum limit requires the measurement of more and more extended defects on 
finer and finer lattices.

For a preliminary exploration we have concentrated on the case of Higgs mass 
$M^*_H=30$ GeV where the phase transition is of strongly first order. At this 
Higgs mass the finite volume effects are numerically small and therefore we 
felt safe to study only one lattice size, $16^3$, at various $\beta_G$. For 
the coarsest lattice we have chosen $\beta_G=\beta^{(0)}_G=8$. Just to be 
specific, we have chosen the lattice spacing corresponding to this gauge 
coupling as the size of topological objects to be examined for a reasonable 
continuum limit. Now we have measured at $\beta_G=k\beta^{(0)}_G$ (for $k=2,3$)
histograms for the densities of extended monopoles and vortices of a size 
(in lattice units) of $k^3$ and $k^2$, respectively. The measurements were 
done near to the pseudocritical $\beta_H$ for the appropriate gauge coupling 
at this lattice size.

The distributions of Nambu monopole and $Z$--vortex densities (for objects of 
corresponding {\it physical} size) are shown in Fig.~4(a) and (b), respectively.
These (normalized) distributions of extended defects have the same two-peak 
structure as elementary ones. With decreasing inverse gauge coupling 
$1/\beta_G$ the metastability becomes weaker if the density of the defects is 
measured. At the smallest $\beta_G=8$, for instance, where we looked for size 
$a$ defects, we could not observe tunnelling between the two phases within 
$10000$ iterations following hot/cold starts. In the Figures, for that 
$\beta_G$, the histograms corresponding to the Higgs phase are omitted since 
only the bin at zero density is populated. The peaks corresponding to the 
symmetric phase shift considerably from elementary objects at $\beta_G=8$
to size-$2$ objects at $\beta_G=16$, but only slightly from this measurement 
to size-$3$ objects at $\beta_G=24$.

We have also checked the percolation of extended vortices. At $\beta_G=16$ on 
a $16^3$ lattice we have found $\beta_{Hc}=0.3391$. Considering all possible 
$k=2$ decimations, we have now analyzed networks of $k=2$ vortices on this 
lattice with respect to percolation. There are $2^3$ different $k=2$ vortex 
configurations per gauge-Higgs configuration which have been obtained at 
$\beta_{H}=0.3390$ and $\beta_H=0.3392$. At $\beta_H=0.3390$ we have found a 
non-zero condensate also for these extended objects, whereas at 
$\beta_H=0.3392$ the condensate is strictly vanishing. We have to admit that 
this result has been obtained on a very small lattice. Nevertheless it is 
indicative for extended vortices percolating at the same temperature.

Similar measurements have to be repeated for defects of different scale
on lattices with different lattice spacings and with bigger, physically
comparable volumes before an understanding of the topological mechanism can be
achieved. Notice that the above results were obtained concentrating, somewhat 
arbitrarily, on topological objects with smallest possible size that could be 
analysed on the coarsest, $\beta_G=8$ lattice. Obviously, the next task is to 
study non-elementary topological objects already on the coarsest lattice. 
This endeavour requires much more computer power in view of the complementing 
measurements that have to be done on finer lattices covering the same volume. 
Since our studies already indicate that (within our lattice construction)
Nambu monopoles and $Z$--vortices as topological excitations of definite
physical size become more and more well-defined in the continuum limit,
it seems to be preferable to dispense with the elementary objects
completely.

The observed, more well-defined behavior is related to the way how the 
fundamental Higgs field is turned into an auxiliary composite field used 
to project out the Abelian fluxes. It seems, that their densities possess a 
sensible continuum limit which can be measured by our lattice prescription.
In contrast to this, elementary monopoles and vortices are still far from the 
continuum limit for defects of corresponding size. A similar problem is 
well-known for instantons in $4D$ lattice gauge theory.

As indicated above, there is another possibility to define extended
defects of any size for a sufficiently fine lattice to start with.
This requires to perform block spin transformations of Higgs and gauge fields
which give a consistent description of the phase transition if seen from
different length scales. In Ref. \cite{stlouis93} two of us have
indicated a practical way how to do this.

\section{Discussion and Conclusions}

We have studied numerically the properties of Nambu monopoles and 
$Z$--vortices which appear as embedded defects in the Standard
model at high temperatures. Our results are obtained in the $3D$
lattice formulation of the $SU(2)$ Higgs sector. We have shown that
these defects are dense in the symmetric phase and that they exist as
a more or less dilute gas in the Higgs phase. Moreover, the percolation
probability of the $Z$--vortex trajectories has been shown to be {\it
the} disorder parameter for the hot electroweak phase. The
$Z$--strings are condensed in the symmetric phase, and the condensate
vanishes in the broken phase. This result has been obtained for the
Higgs mass $M_H^*=30$ GeV at which the phase transition is of
strongly first order. The same is found to be true for the Higgs mass 
$M_H^*=70$ GeV (weakly first order phase transition) where the density 
of monopoles and vortices is clearly discriminating the phases.
Note that in this case at the phase transition the densities in the broken 
phase are {\it not less} than $50$ \% of the corresponding ones in the 
symmetric phase\footnote{Even after the phase transition has ceased to exist 
at larger Higgs masses, preliminary results at $M_H^*=100$ GeV show that the 
condensate of vortices turns to zero at the crossover transition temperature 
\cite{InPreparation}.}.

Comparing the densities of the embedded defects of corresponding
physical size at various values of the gauge coupling
$\beta_G$ gives a first hint that the densities of extended Nambu
monopoles and $Z$--strings as defined in this paper might 
possess a well-defined continuum limit. Further studies are necessary
to understand the dynamics of defects and the spatial structure of
vortices of different physical size near to the transition and in the high 
temperature phase.

The approach of this investigation may be extended to lattice $SU(2)$ Higgs
model in  $D=3+1$. In this case the $Z$--vortex world trajectories
are sweeping out two dimensional world surfaces which are open on the
monopole world lines. According to analytical estimates \cite{Va94}
the $Z$--vortices form a dense network of long strings at temperatures
above a temperature which has been termed Hagedorn temperature $T_H$
in analogy to the dual resonance theory of hadronic strings with its
exponentially rising mass spectrum. One would expect this temperature to be 
of the order of the critical temperature or slightly smaller, $T_H
\sileqq T_c$. Below $T_H$ the network decays into small loops
eventually producing non-zero baryon number~\cite{StringScenario}.
We hope to be able to substantiate this picture by $4D$ lattice studies
in near future.

\section*{Acknowledgements}

M.~N.~Ch. and F.~V.~G. are grateful to M.~I.~Polikarpov and
K.~Selivanov for interesting discussions, E.-M.~I. thanks T.~Suzuki
for useful remarks.

M.~N.~Ch. acknowledges the kind hospitality of the Institute of
Physics of Humboldt University (Berlin) where this project has been
initiated. M.~N.~Ch. and F.~V.~G. were partially supported by the
grants INTAS-96-370, INTAS-RFBR-95-0681, RFBR-96-02-17230a and
RFBR-96-15-96740.

\newpage

\section*{Figures}

\begin{figure*}[!tbh]
\begin{center}
\begin{tabular}{cc}
\hspace{-0.8cm}\epsfxsize=7.1cm
\epsffile{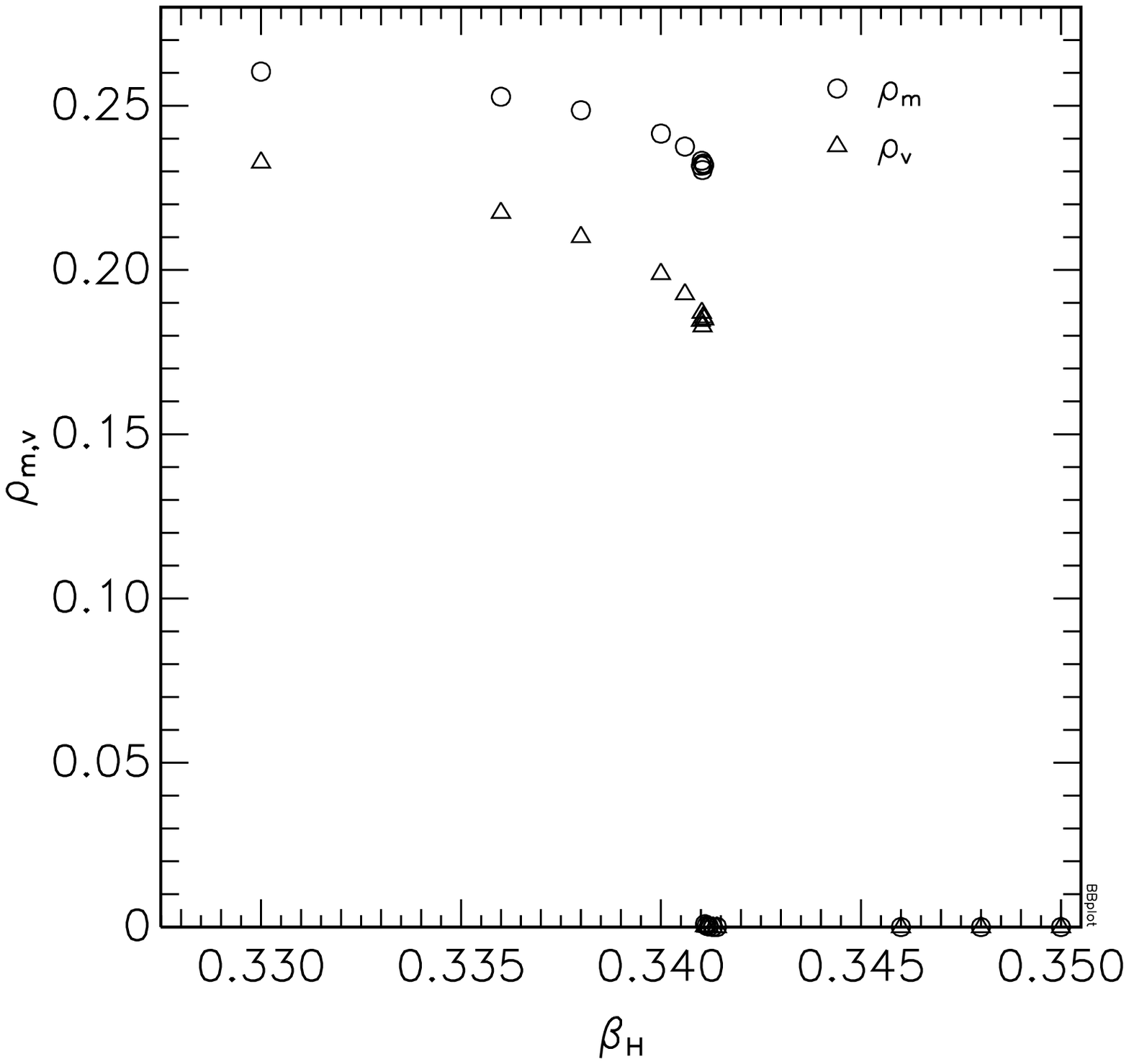} &
\hspace{0.8cm}\epsfxsize=7.1cm
\epsffile{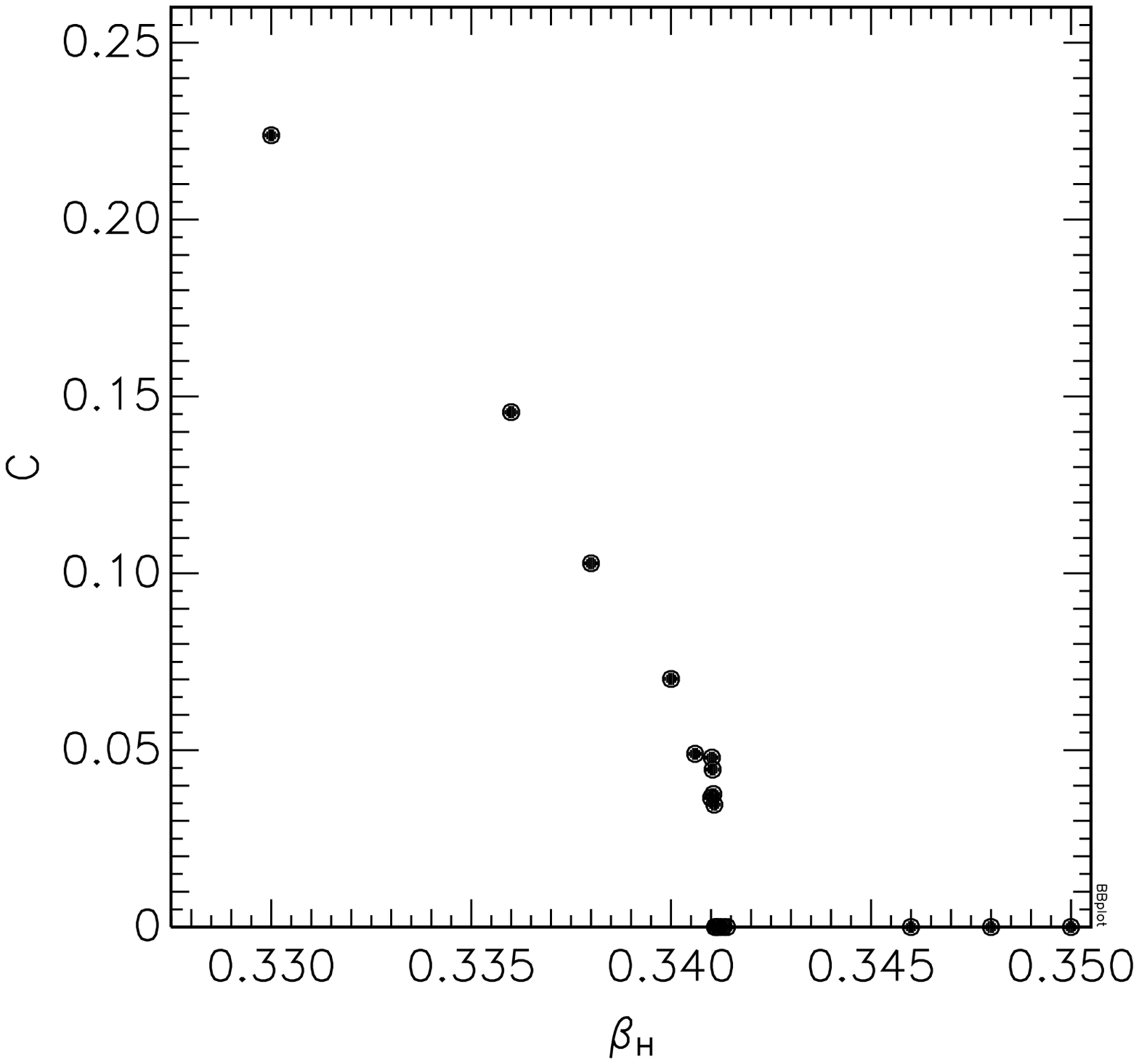} \\
(a) & \hspace{1.5cm}  (b) \\
\end{tabular}
\end{center}
\vspace{-0.5cm}
\caption{(a) Density of  Nambu monopoles $\rho_m$ and
of $Z$--vortices $\rho_v$ {\it vs.} hopping parameter
$\beta_H$ for Higgs mass $M_H^*= 30$~GeV at
gauge coupling $\beta_G = 12$; (b) Percolation probability $C$ of
$Z$--vortex trajectories for the same parameters;
the critical hopping parameter is $\beta_{Hc} \approx
0.3411$.
}
\end{figure*}

\begin{figure*}[!tbh]
\begin{center}
\begin{tabular}{cc}
\hspace{-0.8cm}\epsfxsize=7.1cm
\epsffile{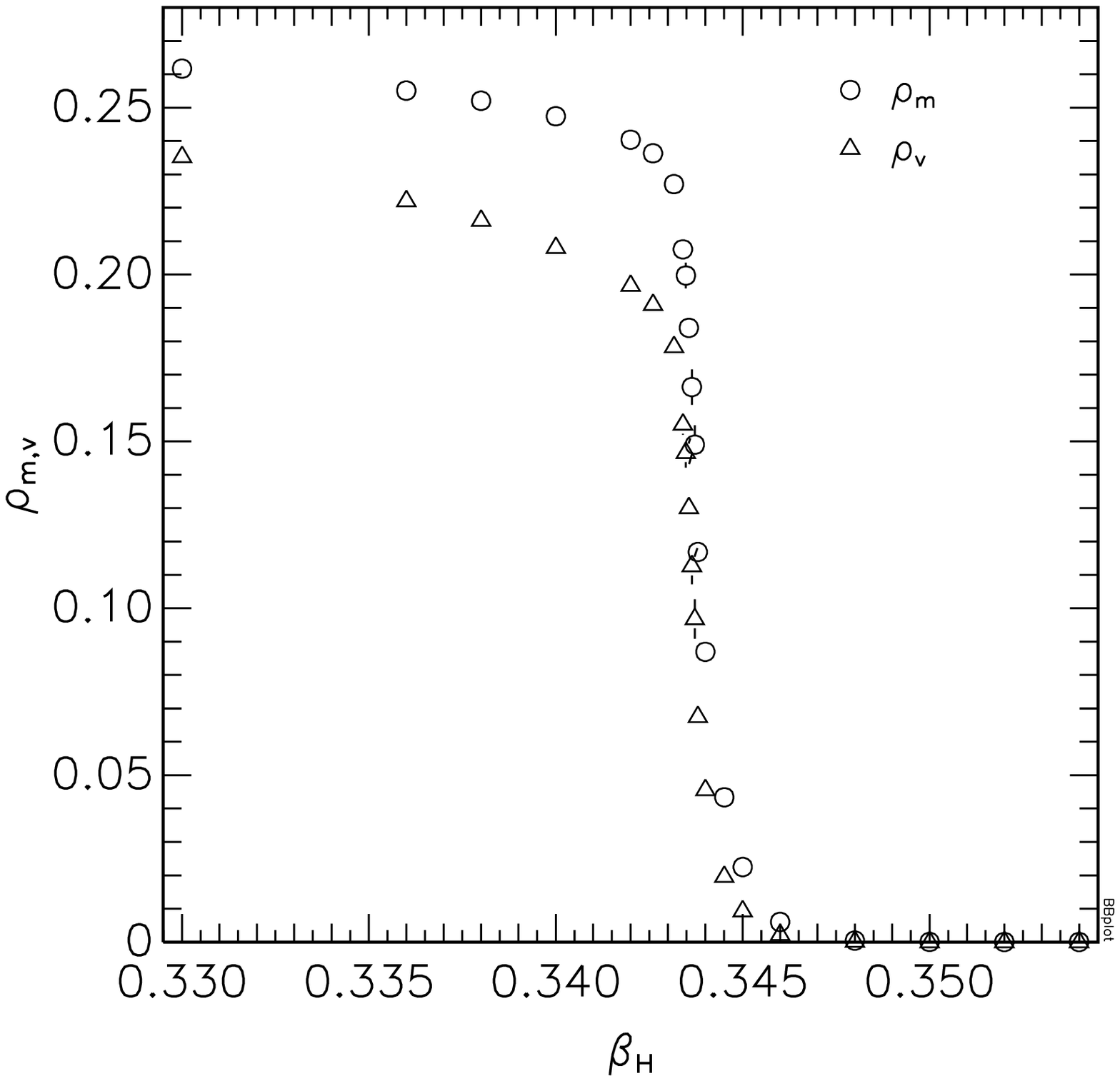} &
\hspace{0.8cm}\epsfxsize=7.1cm
\epsffile{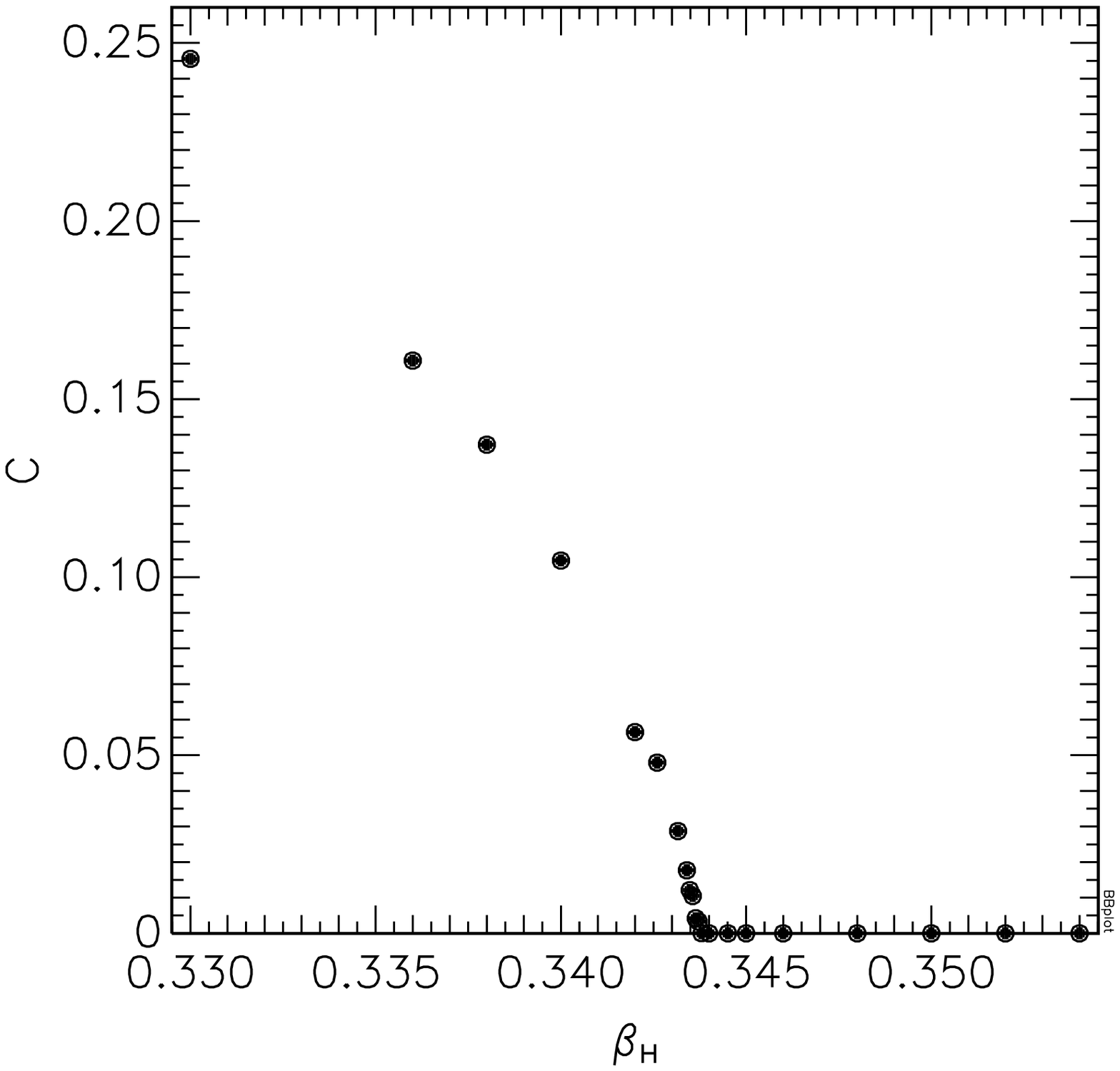} \\
(a) & \hspace{1.5cm}   (b) \\
\end{tabular}
\end{center}
\vspace{-0.5cm}
\caption{Same as in Figure~1 for $M_H^*=70$~GeV;
the critical hopping parameter is $\beta_{Hc} \approx
0.34355$.
}
\end{figure*}

\begin{figure*}[!tbh]
\begin{center}
\begin{tabular}{cc}
\hspace{-0.8cm}\epsfxsize=7.1cm
\epsffile{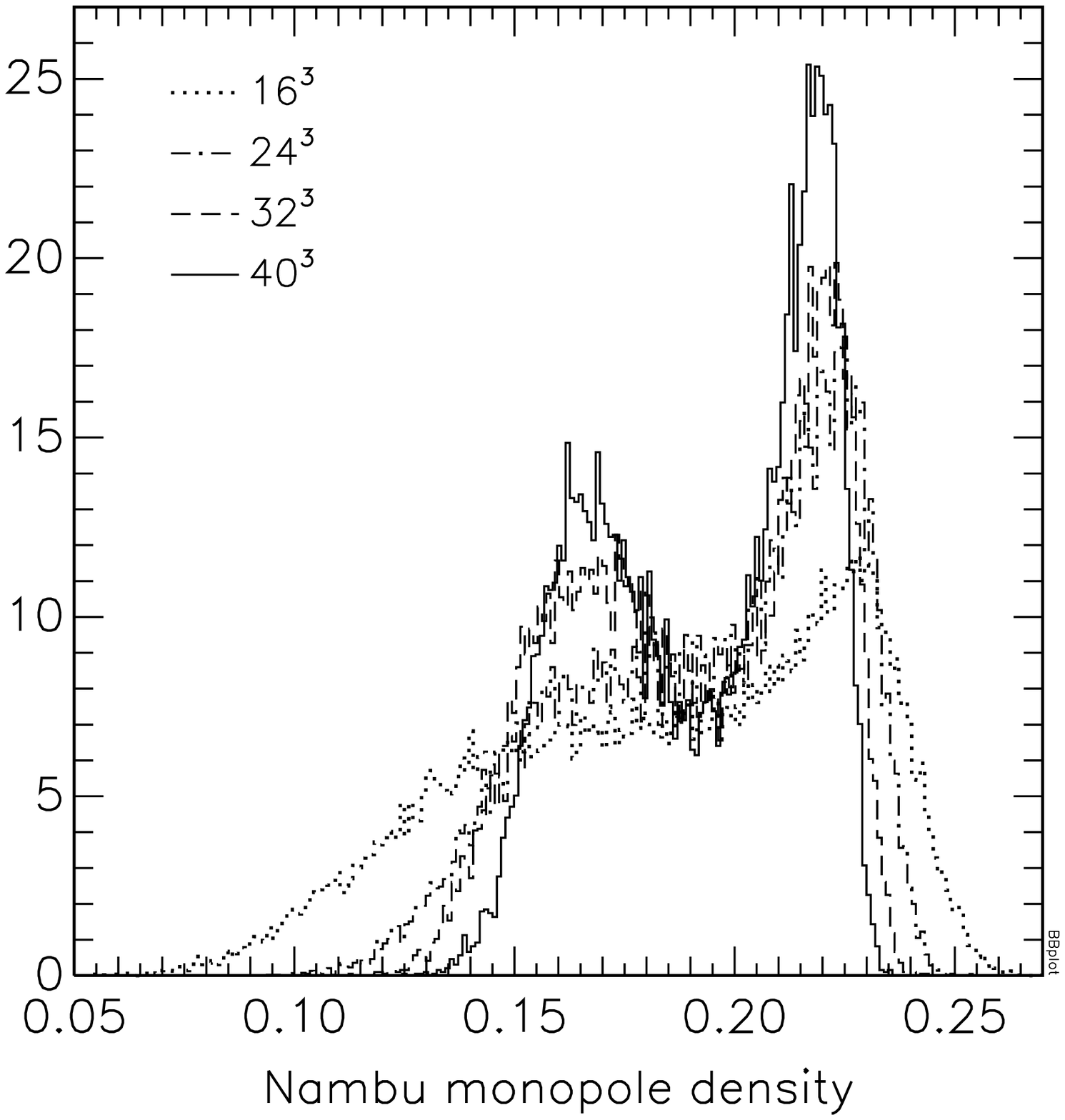} &
\hspace{0.8cm}\epsfxsize=7.1cm
\epsffile{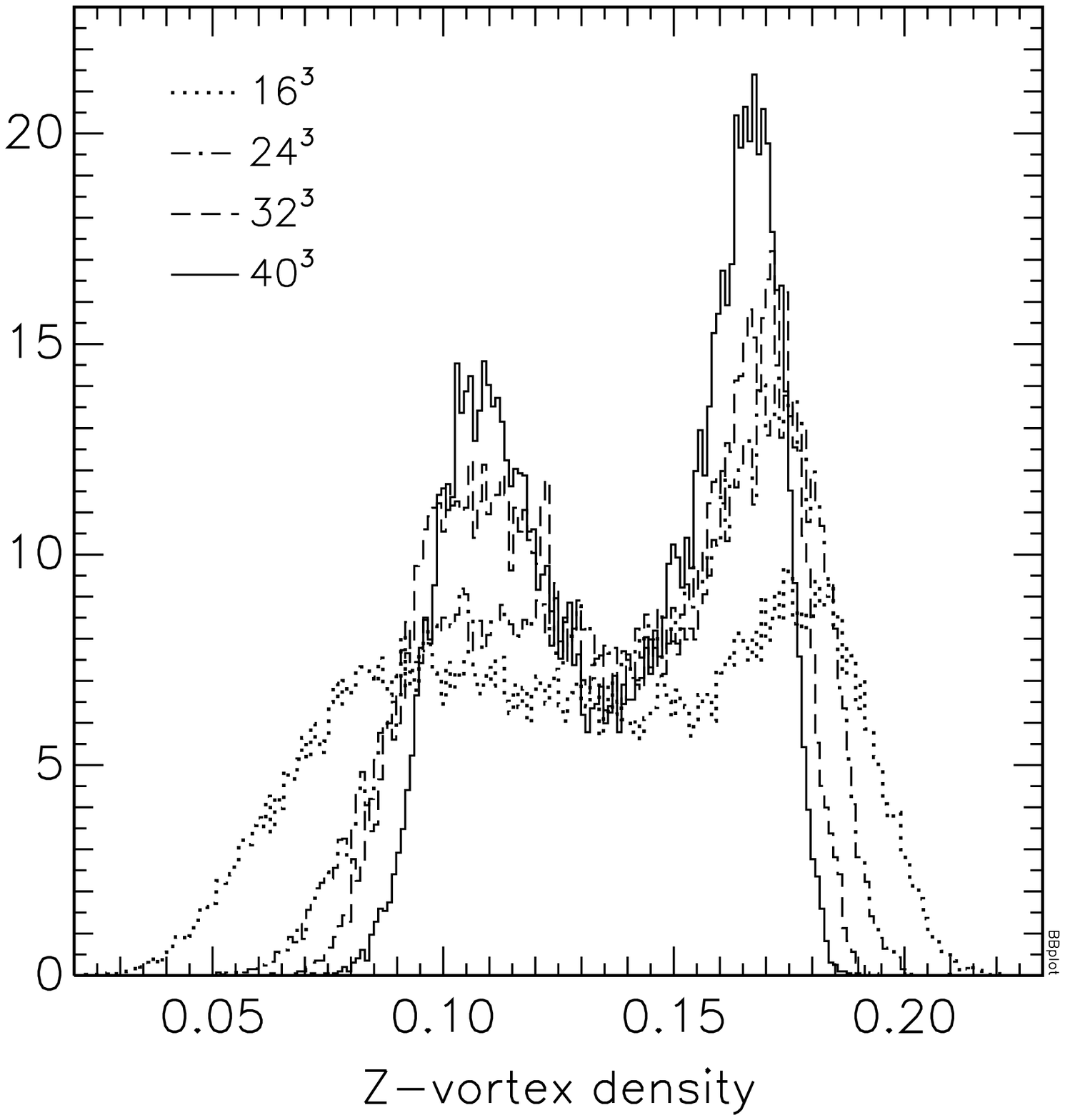} \\
(a) & \hspace{1.5cm}  (b) \\
\end{tabular}
\end{center}
\vspace{-0.5cm}
\caption{Density distributions of Nambu monopoles (a) and of
$Z$--vortices (b) of elementary size at pseudocriticality
for different lattice sizes, $M_H^*=70$~GeV and $\beta_G=12$.}
\end{figure*}

\begin{figure*}[!tbh]
\begin{center}
\begin{tabular}{cc}
\hspace{-0.8cm}\epsfxsize=7.1cm
\epsffile{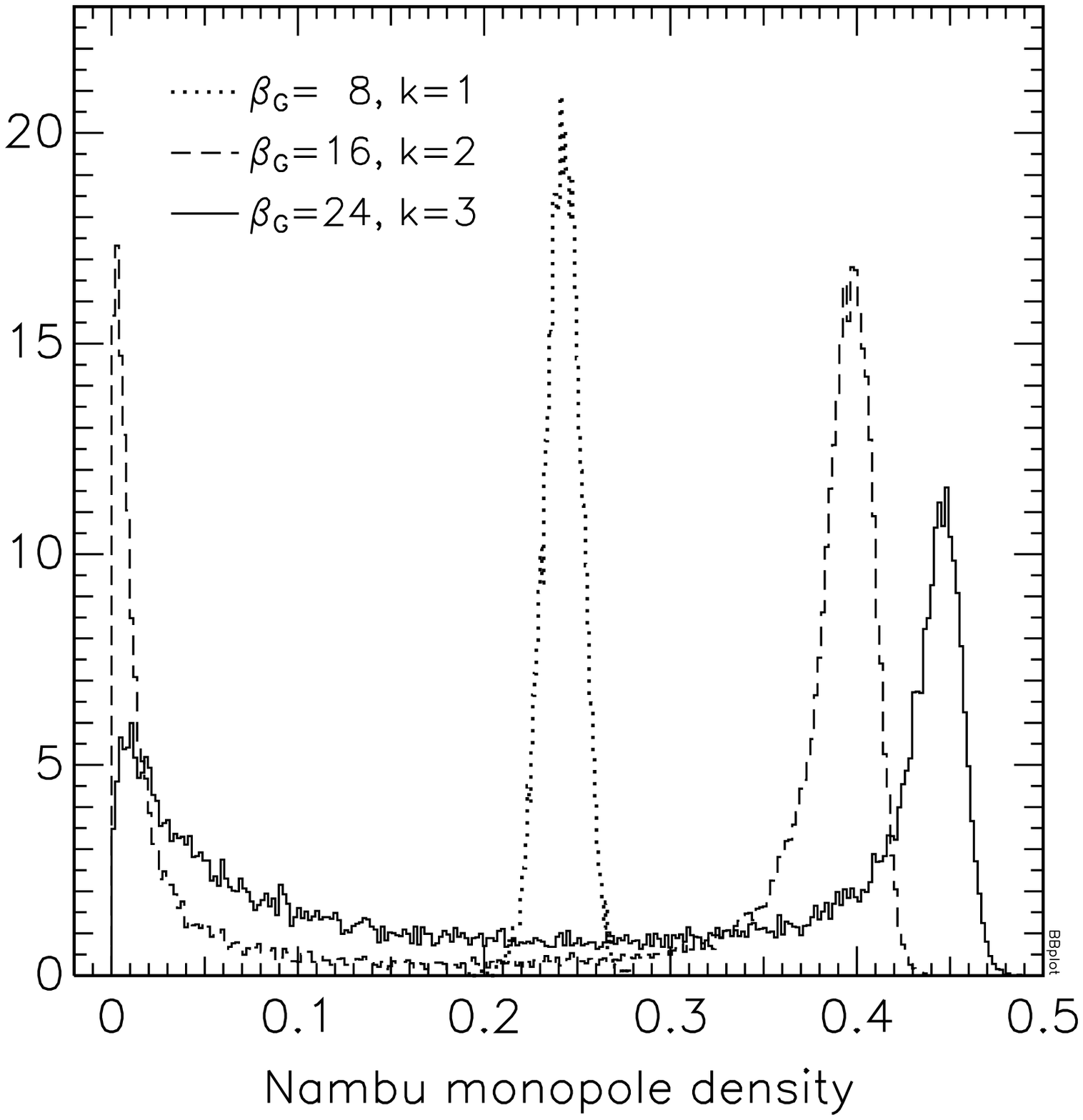} &
\hspace{0.8cm}\epsfxsize=7.1cm
\epsffile{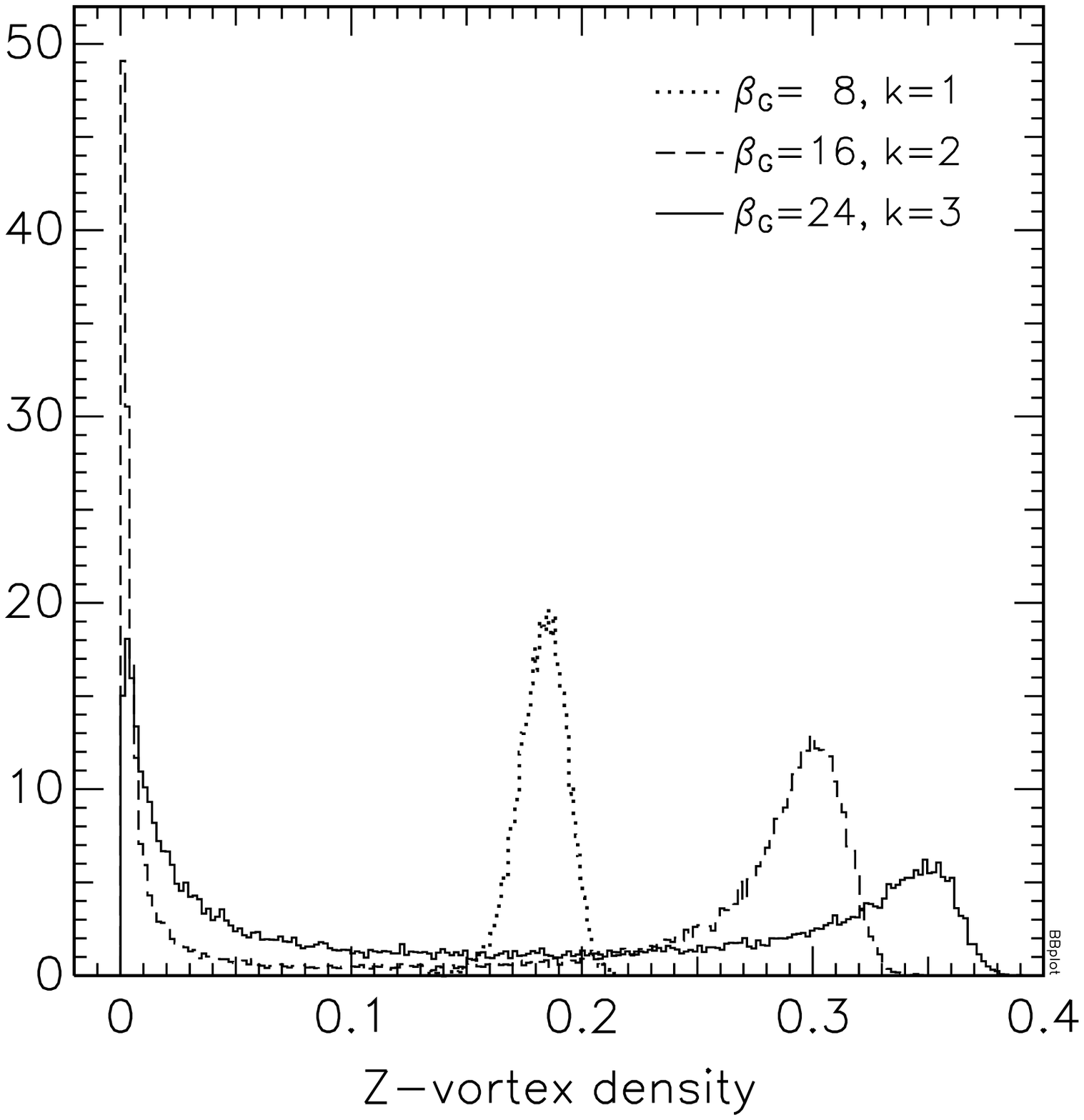} \\
(a) & \hspace{1.5cm}   (b) \\
\end{tabular}
\end{center}
\vspace{-0.5cm}
\caption{Density distributions of Nambu monopoles (a) and of
$Z$--vortices (b) of fixed physical size at pseudocriticality
for different gauge couplings, Higgs mass $M_H^*=30$~GeV and $16^3$ lattice.}
\end{figure*}

\end{document}